\documentstyle[mprocl]{article}

\begin{document}

\title{THE ANALYTIC LANCZOS METHOD}

\author{N. S. WITTE}

\address{School of Physics, University of Melbourne,\\
Parkville Victoria 3052, Australia\\E-mail: nsw@physics.unimelb.edu.au} 

\maketitle\abstracts{
The classical formalism of the Moment Problem
has been combined with a cumulant approach and
applied to the extensive many-body problem.
This has yielded many new exact results for many-body systems in the
thermodynamic limit - for the ground state energy, for excited state
gaps, and for arbitrary ground state averages.
The method applies to any extensive Hamiltonian system, for any
phase or symmetry arising in the model, whether on a lattice or in the
continuum, and for any dimensionality.
The theorems are of a nonperturbative nature with respect to any
couplings occuring in the model. 
}

\section{The Early Development of the Lanczos Algorithm}

The Lanczos algorithm or method has been of interest to physicists
because it is an essentially non-perturbative approach to
physical problems with strong coupling, such as occur in 
the extensive many-body systems of condensed matter physics.
In this method the Hamiltonian is used to generate a sequence
of orthonormal states $ \{ |\psi_{n}\rangle \}_{n=1,2..} $
and Lanczos coefficients 
$ \{ \alpha_n \}_{n=0,1..}, \{ \beta_n \}_{n=1,2..} $,
from a suitably chosen trial state $ |\psi_{0}\rangle $
through the following recurrence
  \begin{equation}
    |\psi_{n+1}\rangle = 
          {1\over \beta_{n+1}}
    [ (\hat{H} - \alpha_{n}) |\psi_{n}\rangle 
                 - \beta_{n} |\psi_{n-1}\rangle ] \ ,
  \label{eq:def-F}
  \end{equation}
so that the Hamiltonian in this new basis is tridiagonal
  \begin{equation}
    T_n = \left(
    \begin{array}{ccccc}
     \alpha_0     &  \beta_1      &            &            &            \\
     \beta_1      &  \alpha_1     & \beta_2    &            &            \\
                  &  \beta_2      & \ddots     &  \ddots    &            \\
                  &               & \ddots     &            &  \beta_n   \\
                  &               &            & \beta_n    &  \alpha_n  
    \end{array}                  \right) \ .
  \label{eq:trid}
  \end{equation}
In the mathematical and statistical arena the Lanczos Process
has been long understood as one manifestation of a body
of intimately connected mathematical subjects, namely -
the Moment Problem\cite{moment-ST,moment-A}, 
Orthogonal Polynomial Systems\cite{moment-ST,moment-A}, 
Pade Approximations\cite{pade-BGM,pade-Br}, 
Analytic Continued Fraction Theory\cite{cf-JT,cf-LW} and 
Krylov Subspace Methods\cite{eigen-S}.
One such equivalence is that the Lanczos process applied up
to $ n_T $ iterations to generate the Lanczos coefficients
$ \{\alpha_{n}\}^{n_T}_{n=0} $, $ \{\beta_{n}\}^{n_T}_{n=1} $
is precisely equivalent to generating the first $ 2n_T+1 $
moments $ \{\mu_{n}\}^{2n_T+1}_{n=0} $ defined by
$ \mu_{n} \equiv \langle \hat{H}^n \rangle $
($ \langle \hat{O}\rangle $ denotes the expectation value with
respect to the trial state).

The traditional use of the Lanczos algorithm has been in a
purely numerical way, that is to say as a numerical
technique for exact diagonalisation of very large matrices
that arise in treating many-body problems in small finite
systems\cite{lanczos-D-94}, or in the treatment of the one-electron
problem in disordered or aperiodic systems, as in the
Recursion Method of Haydock\cite{recur-H}.
The potential of taking the Lanczos algorithm far beyond 
these limitations, into a more powerful, universal formalism
has not been widely appreciated, although some inkling of this
was apparent in the suggestion of Mattis\cite{trid-M-81} concerning 
the exact mapping of the many-body problem onto a one-dimensional
nearest-neighbour model. This idea was explored in some applications
to the Kondo and Wolff models by 
Mancini and Mattis\cite{trid-MM-83,trid-MM-84,trid-MM-85}.
We wish to emphasis to the reader that our approach here is
quite different from that used in the exact diagonalisation studies
of finite systems in two respects - we do not construct a full basis
for a finite system but manipulate basis vectors and coefficients
of an arbitrarily large system analytically and symbolically,
and we perform every iteration exactly and therefore need not
concern ourselves with round-off or loss of orthogonality issues.

While little use or development of the mathematical constructs were 
employed in
exact diagonalisation methods, some of the ideas were used in
other formalisms. Formalisms were developed under the name of the
Recursion Method\cite{recur-H,recur-PW,recur-FH-86,recur-H-90}
or related methods\cite{recur-GCL-73,recur-LG-82,recur-J-85} 
for one-electron problems, but in the last 
analysis every calculation was a numerical evaluation, i.e.
explicit construction of the orthogonal polynomials via 
3-term recurrences and then the continued fraction representation
of the density of states. Some questions were raised concerning 
the generalisation to genuine many-body problems\cite{recur-WO-85}
but this was not realised at the time.
Other formalisms, treating many-body systems and stochastic processes
in the thermodynamic limit,
which arose from this mathematical legacy were the 
Memory Function formalism
\cite{mf-M-65a,mf-M-65b,mf-D-67,mf-GGPS-83,mf-GGP-85},
the Recursion Method\cite{recur-VM} (not to be confused with the 
previous use of the same term) and the Projection Method
\cite{proj-F,proj-BF-88,proj-BWF-89,proj-BF-89,proj-KH-91}. 
However all these methods were applied only formally, that is to say 
the consequences of introducing these tools into the many-body 
problem was not systematically followed through or explored -
the recursion process would be carried out up to a 
finite number of steps and truncated in an ad-hoc manner.
This can be done analytically by hand for the first few steps, but
usually higher steps are calculated on a computer by constructing
an equivalent graphical description of the problem and making
the combinatorial evaluations that arise.
The formalisms are precise and exact in this regard but being 
truncated in this manner they have not converged $ n_T \to \infty $. 
This is a serious issue because while $ n_T $ may be
numerically large, say 20 or 30, one also wants to follow this 
Lanczos convergence with the thermodynamic limit 
$ N \to \infty $, but the problem is that the value of $ n_{opt} $,
to assure convergence to a given accuracy, will scale with $ N $
at best, and may sometimes scale with a higher dependence.

\section{The New Developments}
However it is possible to transcend these limitations in the
process of constructing the mathematical formalism properly
embedded in its physical context. The first key ingredient
is to find a way of incorporating the system size
scaling for the extensive system into the existing formalism from
the outset.
The solution to this is obvious - describe everything in terms
of cumulants, connected moments or semi-invariants\cite{cumulant-K}
$ \{ \nu_n \}_{n=1}^{\infty} $
($ \nu_{n} \equiv \langle \hat{H}^n \rangle_{c} $) instead of moments.
The defining relationship is
  \begin{equation}
       \langle e^{t\hat{H}} \rangle
       = \sum^{\infty}_{n=0} \mu_n {t^n \over n!}
       \equiv \exp\left( \sum^{\infty}_{n=1} \nu_n {t^n \over n!}
             \right) \ ,
  \label{eq:moments}
  \end{equation}
and there exists a unique transformation between the set of
first $ n_T $ cumulants and the set of first $ n_T $ moments.
Cumulants scale with the system size in the following way
  \begin{equation}
     \nu_n = c_n N \ ,\quad
     \nu_n = c_n N + m_n \ ,
  \label{eq:cum}
  \end{equation}
in the ground state sector and other sectors respectively, 
ignoring boundary conditions. 
The coefficients $ c_n $, $ m_n $ are independent of $ N $ and 
functions of coupling constants and other parameters in the trial
state. With this scaling form all information regarding finite-size 
scaling is lost, but it is the simplest approach.
Unconnected moments encapsulate the information about
a system in a very redundant way and which leads to problems of
ill-conditioning. 

Once the above step is taken then many results become quickly
apparent. The first result arises from the substitution of
the cumulant Eq.~(\ref{eq:cum}) into the explicit expression 
for the moments, and then into the Lanczos coefficients,
and expanding the resulting forms in a large $ N $ expansion -
and the result is a remarkably simple and perfectly universal
expansion
 - the ``Plaquette Expansion''\cite{pexp-L89,pexp-1st-H,pexp-proof-WH}
 - as a function of an arbitrary Lanczos iteration number $ n $
\begin{equation}
  { \alpha_{n} \over N }
  =  c_1 + n \left[ {c_{3}\over c_{2}} \right] {1\over N} 
     + {1\over 2}n(n-1)
    \left[ {3 c_3^3\!-\!4 c_2 c_3 c_4\!+\!c_2^2 c_5 \over 2 c_2^4} 
    \right] {1\over N^2}
     + \dots \ ,
\label{eq:alpha_PE}
\end{equation}
and
\begin{eqnarray}
   { \beta_{n}^{2} \over N^2 }
 & = & n c_{2} {1\over N}
          + {1\over 2}n(n-1)
          \left[ {c_{2}c_{4}\!-\!c_{3}^{2}\over c_{2}^{2}} \right]
          {1\over N^2}
 \nonumber \\
 && \quad + {1\over 6}n(n-1)(n-2) 
    \left[ {-12 c_{3}^4\!+\!21 c_{2} c_{3}^2 c_{4}
              \!-\!4 c_{2}^2 c_{4}^2\!-\!6 c_{2}^2 c_{3} c_{5}
                \!+\!c_{2}^3 c_{6} \over 2 c_{2}^5}
    \right] {1\over N^3}
          + \dots \ . 
\label{eq:beta_PE}
\end{eqnarray}
It can be shown\cite{alm-mops-W} that just retaining the first 
terms in each coefficient can be interpreted as a manifestation of 
the Central Limit Theorem and a description in terms of Gaussian 
fluctuations, while retaining the first two terms in each describes
dynamical processes governed by the Binomial distribution and 
all related ones. In this way many of the distributions of 
statistics arise naturally in such an expansion, and in fact simple
interacting physical models can be found which are exactly 
represented by a finite number of terms in this 
expansion\cite{alm-mops-W}.

It should be noted that in each term of the above the degree of
the polynomial in $ n $ is the same as the inverse power of
$ N $ so that the following limit $ n,N \to \infty $ exists
at fixed $ s\equiv n/N $. Although the above is just a
Taylor series expansion in $ 1/N $ we conjecture that the 
exact Lanczos coefficients exhibit the following general
confluence
  \begin{equation}
     \alpha_n(N)   \stackrel{n,N \to \infty}{\rightarrow} 
                   N\alpha(s) \ ,\quad
     \beta^2_n(N)  \stackrel{n,N \to \infty}{\rightarrow} 
                   N^2\beta^2(s) \ .
  \label{eq:confl}
  \end{equation}
In the next step if one defines the spectral envelope functions
  \begin{eqnarray}
    e_n(N) & = &
    1/2\left\{ \alpha_{n}\!+\!\alpha_{n-1}
        - \sqrt{(\alpha_{n}\!-\!\alpha_{n-1})^2\!+\!16\beta^2_n}
       \right\} \ ,
           \nonumber \\
           & \stackrel{n,N \to \infty}{\rightarrow} 
           & e(s) = \alpha(s)-2\beta(s) \ ,
  \label{eq:spec_env}
  \end{eqnarray}
then one can employ theorems on bounds to the extremal zeros of
Orthogonal Polynomials to arrive at an exact theorem for the 
ground state energy\cite{pexp-exactgse-HW}
  \begin{equation}
      \epsilon_0 = \inf_{s}[e(s)] \ ,
  \label{eq:GSE}
  \end{equation}
and if this occurs at a finite point it is denoted $ s_0 $.
This result constitutes an exact diagonalisation of the 
many-body problem in the thermodynamic limit, as the formalism 
expresses results in terms of the tridiagonal matrix elements,
or Lanczos coefficients. From this it is a simple step to finding 
the ground state average for an arbitrary operator 
$ \hat{O} $\cite{pexp-2dhm-WHW}
  \begin{equation}
      \langle \hat{O} \rangle = \left[     
                        \delta^O\!\alpha(s)-
                      { \delta^O\!\beta^2(s) \over \beta(s) } 
                          \right]_{s_0} \ ,
  \label{eq:GSA}
  \end{equation}
where the operator Lanczos coefficients are constructed from the 
operator cumulants
  \begin{equation}
    \delta^O\!\nu_{n+1} = \sum^{n}_{k=0} 
    \langle \hat{H}^{n-k} \hat{O} \hat{H}^k \rangle_{c}
    \to \delta^O\!\alpha,\delta^O\!\beta^2 \ ,
  \label{eq:GSA_cum}
  \end{equation}
in the same manner as the pure Lanczos coefficients.
The excited state gap, between the ground state and an excited
state in another sector, 
is just the difference between two ground state energy densities, 
$ \epsilon_1-\epsilon_0 = \Delta\epsilon/N $
and is thus\cite{pexp-tgap-HWW}
  \begin{equation}
      \Delta\epsilon =  \left[     
                        \delta^G\!\alpha(s)-
                      { \delta^G\!\beta^2(s) \over \beta(s) }
                         \right]_{s_0} \ ,
  \label{eq:GAP_T}
  \end{equation}
where the gap Lanczos coefficients are constructed from the gap
cumulants (constructed using a trial state with the excited state
quantum numbers)
  \begin{equation}
    \nu_{n} = c_n N + \delta^G\! c_n
    \to \delta^G\!\alpha,\delta^G\!\beta^2 \ .
  \label{eq:GAP_cum}
  \end{equation}
For the excited state gap in the same sector, the following peeling
theorem holds\cite{pexp-sgap-WH}
  \begin{equation}
      \Delta\epsilon = 2 \lim_{n,N \to \infty} N \left[     
                               e(s) - e_n(N) \right]_{s_0} \ .
  \label{eq:GAP_S}
  \end{equation}
In its application to non-integrable models the above expansion,
Eq.(\ref{eq:alpha_PE},\ref{eq:beta_PE}), is generated from a finite
set of low order cumulants and then 
truncated at some finite order and the above theorems applied without 
the need for any extrapolation. Some examples where this has been
successfully employed are the 1 and 2-dimensional Heisenberg 
models\cite{pexp-1dhm-WH,pexp-2dhm-WHW} and lattice gauge 
models\cite{pexp-su2-H,pexp-qcd-H}.
There are also examples of this method used in an essentially exact 
manner, namely for a 1-dimensional solvable spin model with a phase 
transition at $ T=0 $\cite{pexp-itf-WS}, 
where the convergence properties of the method have been examined. 

\section{Summary}
As we have seen the Analytic Lanczos Method is an important stage
in the development of Lanczos methods in the extensive many-body
problem. Amongst its virtues are that is general purpose - it works
for any Hamiltonian, lattice or continuum, quantum mechanical 
or classical, in all dimensions D, for any phase or symmetry of the
model, it is non-perturbative in couplings, 
it works exactly in the thermodynamic limit $ N\to \infty $
and it applies to ground state or $ T>0 $ properties.
It is accurate and systematic in that there is a development in 
successive orders so that some control of the errors can be made.
It has a flexible implementation in that the treatment can be
either analytic, semi-analytic or numerical depending on the 
degree of integrability of the model at hand, that one is free to
choose the trial state, within very general limits relating to the
symmetry of the target state, and that one
can combine it with other methods, e.g. variational, mean-field, \dots.

\section*{Acknowledgements}
The author would like to acknowledge support from the Australian
Research Council in preparing this report and for many discussions
with L.C.L. Hollenberg and other members of the Melbourne High
Energy Physics Group.

\vspace*{-9pt}

\section*{References}
\bibliographystyle{aip}    % for BibTeX - sorted numerical labels by
                             % order of first citation.
\bibliography{moment,pexp,texp}
\eject

% \begin{figure}
% \rule{5cm}{0.2mm}\hfill\rule{5cm}{0.2mm}
% \vskip 2.5cm
% \rule{5cm}{0.2mm}\hfill\rule{5cm}{0.2mm}
% \psfig{figure=filename.ps,height=1.5in}
% \caption{A generalized cactus tree: the confluent
% transfer-matrix $S$ transforms the state function $f(x)$ and
% $f(z)$ into $f(x)$.  \label{fig:radish}}
% \end{figure}

% \vspace*{-2pt}

\end{document}